\documentclass[prl,aps,preprint,showpacs,showkeys]{revtex4}
\begin{document}
\setcounter{page}{1}
\date{\today}
\title{Varying critical percolation exponents on a multifractal support}

\author{  J. E. Freitas, G. Corso$\dagger$,  and L. S. Lucena }

\affiliation{ International Center for Complex Systems and
 Departamento de F{\'\i}sica Te\'orica e Experimental,
 Universidade Federal do Rio Grande do Norte, Campus Universit\'ario
 59078 970, Natal, RN, Brazil.}

\affiliation{ $\dagger$ Departamento de Biof{\'\i}sica,
    Centro de Bioci\^encias,
Universidade Federal do Rio Grande do Norte, Campus Universit\'ario
 59072 970, Natal, RN, Brazil.}

\begin{abstract}
We study percolation as a critical phenomenon on a multifractal
support. The scaling exponents of the
the infinite cluster size ($\beta$ exponent) and the
fractal dimension of the percolation cluster ($d_f$) are quantities
that  seem do not depend on local anisotropies. These
two  quantities have the same value as in the standard percolation
 in regular bidimensional lattices.
On the other side, the scaling of the  correlation length ($\nu$
exponent) unfolds  new universality classes
  due to the local anisotropy of the critical percolation
cluster.  We use two  critical exponents $\nu$ according to
the percolation criterion for crossing the lattice in either direction
or in both directions. Moreover $\nu$ is related to a parameter
that characterizes the stretching of the blocks
forming the tilling of the multifractal.

\end{abstract}



\maketitle

\section{1 - Introduction}

One remarkable feature of critical phenomena is the concept of
universality, that means, the fact that critical exponents are
independent of the particular type of lattice and of nature of
the physical system.  Transitions in systems so diverse as gases,
magnets or metalic alloys, can be characterized by the same
critical indices. In some sense this can be explained as
a consequence of the equivalence of the local simmetry of these
different systems \cite{ShangMa}.
Non universal behavior is a rare event
observed only in special situations and non trivial models.

In magnetic systems the critical exponents, in
general, depend only on the dimension of the space and on the
dimension of the spins. The variation of critical exponents for fixed dimension of the
lattice and fixed dimension of the spin is found in non trivial cases
in which there is some types of competing interactions, frustration or
symmetry breaking \cite{Domb&Green, Stanley}.

In this work we report results obtained for a complex model which
is universal with respect to some exponents but that shows
non-universality with respect to the correlation length exponent $\nu$.
We deal with percolation phenomena in a multifractal support immersed in
a 2D space. It has been shown that percolation corresponds to the $q=1$
state of Potts model \cite{Wu}. In this perspective we expect universal behavior for
percolation in $2$ dimensions, for any kind of lattice. The
multifractal support we have proposed has a new local lattice structure: the
local connectivity and the local anisotropy
vary from region to region. The strength of the anisotropy change of the
multifractal support is related to a single parameter $\rho$. The
exponent $\nu$ of the critical percolation cluster varies with $\rho$
showing an accentuated sensibility to the local anisotropy of the
support structure. On the other side, the percolation cluster is a fractal
with a fractal dimension, $d_f$,
that is identical to the one of the standard percolation cluster. The
same occurs with the critical exponent $\beta$. That means, if we take
into consideration only $d_f$ and $\beta$ we could say that the
universality class holds for this problem.

A motivation for this work also comes from the problem of direct percolation
in which the anisotropy of the model is responsible for anisotropic clusters and for
different correlation lengths in parallel and perpendicular directions.
In this work we observe the build up of percolation clusters that
are very elongated for small values of $\rho$, showing anisotropic
structures. Do these local anisotropic properties affect the
critical exponents?

In the reference \cite{last,lim} an  object, $Q_{mf}$,
is developed to study percolation in multifractal systems.
This object is a natural generalization of the square lattice in
which it is possible to estimate analytically the full spectrum
of fractal dimensions. Besides it exhibits a rich distribution
of area among the blocks and a non-trivial topology, the
number of neighbors varies along the object. Moreover its
geometric qualities and  algorithmic simplicity, it is
straightforward to study the percolating properties of $Q_{mf}$.

In a previous work \cite{last} we have introduced the multifractal,
analytically determined its spectrum of fractal dimensions, analyzed some
properties of the histogram of percolating lattices versus
occupation probability, and estimated the fractal dimension of
the percolation cluster for several parameters of $Q_{mf}$.
In a further work \cite{lim} we have performed a comprehensive study of
the determination of the percolation threshold, $p_c$. In
particular we have explored the relation between $p_c$ and the
topologic properties of $Q_{mf}$. In this work we  deal
with the problem of the class of universality of $Q_{mf}$ and
its relation with symmetry characterization.

The multifractal object we develop, $Q_{mf}$,
is an intuitive generalization of the square lattice. Suppose that in the
construction of the square
lattice we use the following algorithm: take a square
of size $L$ and
cut it symmetrically with  vertical and  horizontal lines.
Repeat this process
$n$-times; at the $n^{th}$ step we have a regular lattice
with $2^n \times 2^n$
cells. The setup algorithm of $Q_{mf}$ is quite similar,
the main difference
is that we do not  cut the square in a symmetric way.
In the next section we explain in detail this algorithm.

\section{2 - The Multifractal support}

In this section we present the multifractal object
$Q_{mf}$ in a double perspective. Initially we introduce the
object $Q_{mf}$ using its growing algorithm. In a second moment we show
how percolation properties are studied in such a non-trivial
system.

\subsection{2-1 The generating algorithm of $Q_{mf}$}

 We start with a square of linear size $L$ and a  parameter
$0<\rho = \frac{s}{r} <1$, for integers $r$ and $s$.
The first step, $n=1$, consists of two sections of
the square: a vertical and an horizontal.
Initially the square is cut in two pieces of
area $\frac{s}{r+s}$ and $\frac{r}{s+r}$ by a vertical line
(we use $L^2$ area units).
This process is shown in figure \ref{fig1} (a), where we use
$\rho = \frac{s}{r} = \frac{2}{3}$.
 The horizontal cut is shown in figure \ref{fig1} (b).
Note that we use the same  $\rho$.
The first partition of the square generates
four rectangular blocks: the largest one of area $(\frac{r}{s+r})^2$,
two  of area $\frac{r \> s }{(r+s)^2}$ and the
smallest one of area $(\frac{s}{r+s})^2$.

The second step, $n=2$, is shown in figure \ref{fig1} (c) and  (d).
The same process of vertical and horizontal
sections is repeated as in step $1$ inside each block. In this way
$Q_{mf}$ is a self-affine object.
 As observed in the figure at end of step $n=2$ there are $2^4$ blocks.
Generically we have after the $n^{th}$-step $2^{2 n}$ blocks.
The partition process produces a set of blocks with a variety of areas.
 We call a set of all elements with the same area as
a $k$-set. In the case $n=2$ there are five $k$-sets.

 At the $n^{th}$-step of the algorithm, the
partition of the area $A$ of the square in   $L^2$ units
 follows the binomial rule:
\begin{equation}
    A \> = \> \sum_{k=0}^n \> C_k^n \> \left(\frac{s}{s+r}\right)^k \left(\frac{r}{s+r}\right)^{n-k}
                = \> \left(\frac{r+s}{r+s}\right)^n = 1.
\label{bino2}
\end{equation}
The number of elements of a $k$-set is $C_k^n$.

In reference \cite{last} we see that as $n \rightarrow \infty$ each  $k$-set
determines a monofractal whose dimension is
$d_k = lim_{n \rightarrow \infty} \frac{log \> C_n^k \>
\> s^k \>r^{(n-k)} }{log \> (s+r)^\frac{n}{2} }$. In the same  limit
 $n \rightarrow \infty$ the ensemble of all
$k$-sets engenders the actual multifractal $Q_{mf}$.

In figure \ref{fig2} (a) we show $Q_{mf}$ ,
as in figure \ref{fig1} ($\rho=\frac{2}{3}$), for $n=8$. We use
the following code color: blocks of equal area (same $k$-set)
share the same color.  Figure \ref{fig2} (b)
is a zoom of the inner square part of
\ref{fig2} (a). We can observe in this picture
that it shows a slight anisotropy compared with
the square lattice. The anisotropy
in the figure is not very accentuated because $\rho=\frac{2}{3}$ is close to
the square lattice case, $\rho=1$.

In the construction process of $Q_{mf}$ we see that as $r>>s$ (or
$\rho \rightarrow 0$) the blocks became more stretched in one direction
than other.
 In this way $\rho$ is a measure of the stretching of the blocks. This
property will reflect in the anisotropy of the percolation
cluster as we shall see in the simulations.

\subsection{2-2 The percolation in  $Q_{mf}$}

The main subject of this work is the study of the percolation properties of
$Q_{mf}$. To perform such a task we develop a percolation algorithm.
The percolation algorithm of   $Q_{mf}$
starts mapping this object into the
square lattice. The square lattice should be large enough that
each line segment of $ Q_{mf}$
coincides with a line of the lattice, this condition
imposes that $\rho$ is a rational number.
 In this way all blocks of the multifractal are
composed by a finite number of cells of the square lattice.
To explain the percolation algorithm we suppose that  $Q_{mf}$
construction is at step $n$.
We proceed the percolation algorithm by choosing at
random one among the $2^{2 n}$ blocks of
$ Q_{mf}$ independent of its size. Once a block is chosen all the  cells
in the square lattice corresponding to this block are considered
occupied. Each time a block of $ Q_{mf}$ is chosen
the algorithm check if the occupied cells at the underlying square
lattice are connected in such a way to form an $infinite$ percolation
cluster. The algorithm to check percolation is similar to the one used in
 \cite{ziff,bp2,freitas0,freitas2}.

Figure \ref{fig3} shows the percolating cluster of $Q_{mf}$ for
 $\rho=\frac{1}{3}$ and a particular realization.
Others $\rho$ show a similar aspect, but the
blocks became more stretched  as $\rho \rightarrow 0$. In this limit
 the multifractal turns more  anisotropic. If we compare this
percolating cluster with the percolating cluster of square lattice \cite{stauffer} we
observe a strong anisotropy in the present case. Such anisotropy
is observed also in the scaling of the correlation length, $\xi$, as we
see in the next section.

Following the literature \cite{stauffer} we call $p$ the
density of occupation of
a lattice. $R_L$  is the probability that for a site occupation $p$ there exists a
contiguous cluster of occupied sites which cross completely  the square lattice
of size $L$.
$p_c$ is the density of occupation at the percolation threshold.
There are several ways \cite{ziff} to define $R_L$, we use two of them:
$R_L^e$  is the probability that there exits a cluster crossing
either the horizontal or the vertical direction, and $R_L^b$ is the
probability that there exits a cluster crossing around
both directions. At the limit
of infinite lattice size $R_L^e$ and $R_L^b$ converge to a common value
in the case of the square lattice.

\section{3 - Numerical Results}

In this section we deal with the estimation of critical
exponents and fractal dimensions of  $Q_{mf}$.
Before the numerics we warn to a specific point in this process.
Suppose the multifractal construction is at step $n$, the underlying
square lattice has the number of cells $2^n \times 2^n$, and the
number of  blocks  is $2^{2n}$, values that are
independent of $\rho$. The size of the underlying lattice, however,
is $L=(r+s)^n$, a function of $\rho$. Because of this point we
have to take care in the effect of finite size  properties
of the multifractal.

A comprehensive study of the percolation threshold, $p_c$,
 of $Q_{mf}$ is performed in  \cite{lim}, we resume this point
 in the following. The way we  find the best
estimative of $p_c$ is to use, for several $L$,
the average value of $R_L^e$ and $R_L^b$. The results for
specific values of $\rho$ are summarized in Table $I$. The
results point to a
value of $p_c$ which slightly increases with $\rho$ and
present a strong discontinuity at $\rho = 1$ (the square lattice).

\subsection{3-1 Fractal dimension of the percolating cluster}

The estimation of the fractal dimension of the percolating cluster, $d_f$,
is performed using the definition:
\begin{equation}
               d_f  = lim_{L \rightarrow \infty} \frac{ln (M)}
               {ln ( L)}.
\label{df}
\end{equation}
Where $M$ is the mass of the percolating cluster and
$L$ the lattice size. We use in our simulations values of
$L$ that correspond to $5<n<10$. The results for several
$\rho$ are shown in Table $I$. The exact value of $d_f$ in two dimensions
is $d_f=\frac{91}{48}=1.89583$. The values obtained are close to $2\%$ of
the exact value and are not correlated to $\rho$.
However, this fact is not enough to conclude that $Q_{mf}$ is
in the same class of universality of two dimensional standard percolation.
In the next paragraphs we shall return to this point.

\subsection{3-2 The critical exponent $\beta$}

Percolation in bidimensional spaces  shows  critical phenomenon close
to the critical point $p_c$.
The critical exponent $\beta$  is defined
from the relation:

\begin{equation}
                 R_L \sim \left( p_c(L) - p_c \right)^{\beta},
\label{beta}
\end{equation}
where $p_c$ is the exact occupation probability value in contrast to
 $p_c(L)$ which is its finite size value. The power-law (\ref{beta})
 is verified for $p_c(L)$ obtained either from $R_L^e$ or $R_L^b$. In the first case
the formation of a spanning cluster and the power-law are at $p>p_c$. Otherwise,
if we use $R_L^b$, we have to search for a power-law in the vicinity of
the critical value for $p<p_c$. We use in our estimation $R_L = R_L^e$ which
shows smaller fluctuations than $R_L^b$.
As the numerical estimation of $\beta$ is based in equation (\ref{beta}),
$R_L$ is a key element of the analysis. For $Q_{mf}$ the probability
 $R_L$ is not a well
behaved function of $p$ for low $L$ \cite{last}.
Actually, $R_L$  can show,
depending on $\rho$, an inflection point at $p_c$ in this regime. In the
case where $L \rightarrow \infty$ the scaling of $\left( p_c(L) -p_c \right)$
recovers the power-law behavior. In this regime we find the same $\beta$
 of two dimension standard spaces,
$\beta=\frac{5}{36}=0.13888$. We check in our simulations that for $n=9$,
 $\beta$ is around $5\%$ of the exact
value of standard percolation. A set of values of $\beta$ are shown in Table $I$.

Table $I$ summarize the estimated $p_c$, $d_f$, and $\beta$ for
several $\rho$. The values of $d_f$, and $\beta$ fluctuate
around the theoretical value of the standard percolation and
do not show any correlation with $\rho$.
The main conclusion we take from the data is that
$d_f$ and $\beta$ do not depend on $\rho$.

\centerline{\it Table I}
\begin{tabbing}
  $(s,r)$      \hspace{0.6cm}        \= (1,1) \hspace{0.7cm} \= (4,3) \hspace{0.7cm}\=      (3,2) \hspace{0.7cm} \= (2,1) \hspace{0.7cm} \= (5,2)     \hspace{0.7cm} \=(3,1)      \hspace{0.7cm} \= (4,1)   \hspace{0.7cm} \= (5,1)\\
\hspace{0.2cm} $p_{c}$  \hspace{0.6cm} \= 0.5929 \hspace{0.4cm} \= 0.5253  \hspace{0.4cm} \= 0.5267  \hspace{0.4cm} \= 0.5262   \hspace{0.4cm} \=  0.5260   \hspace{0.4cm} \= 0.5261  \hspace{0.4cm} \= 0.5254   \hspace{0.5cm} \= 0.5253 \\
\hspace{0.2cm} $ d_f$   \hspace{0.6cm} \= 1.895 \hspace{0.6cm} \= 1.892 \hspace{0.7cm} \= 1.890  \hspace{0.6cm}  \= 1.900 \hspace{0.7cm} \= 1.891    \hspace{0.6cm} \= 1.911  \hspace{0.7cm} \= 1.902   \hspace{0.6cm} \= 1.929 \\
\hspace{0.2cm} $ \beta$   \hspace{0.7cm} \=0.127 \hspace{0.6cm} \= 0.135 \hspace{0.7cm} \= 0.141   \hspace{0.6cm}  \= 0.128  \hspace{0.7cm} \= 0.131         \hspace{0.6cm} \= 0.140    \hspace{0.7cm} \= 0.141   \hspace{0.6cm} \= 0.118 \\
 \end{tabbing}

\subsection{3-3 The critical exponent $\nu$}

An important quantity in percolation theory is
the  correlation length, $\xi$. It diverges at the critical point $p=p_c$
with a critical exponent  $\nu$ and it is related to the size of
the percolation cluster.  In a finite size system, near the critical point,
we use this fact to approximate $\xi$ by the system length $L$. The exponent $\nu$ is
defined from the scaling equation:
\begin{equation}
          \xi \sim L \sim \left( p_c(L) -p_c \right)^{-\nu}.
\label{nu}
\end{equation}
In order to estimate numerically $\nu$ we plot
$ln  \left( p_c(L) -p_c \right)$ against $ln (L)$. The slope of this curve is
 $\nu$. Figure \ref{fig4} illustrates this process for $\rho=\frac{1}{3}$ and
the case where $p_c$ is obtained from $R_L^e$.
It is indicated in the figure the
 linear regression equation of the data.  $\nu$ can also give some
 hint about the symmetry of the percolation cluster. In other words,
 $\nu$ informs about the isotropy of the percolation cluster, or the
 dependence of $\xi$ with some direction.

We emphasize that $Q_{mf}$ is not an isotropic object and as a
consequence the percolating cluster is also anisotropic. In this
way there is no reason for the percolating cluster to be
associated to an unique scaling relationship as in equation (\ref{nu}).
We note that $R_L^e$ is related to the
existence of a spanning cluster in either direction,
and $R_L^b$ in  both directions. Actually $R_L^b$ measures an
average of the crossing probability over both directions. This average
over both directions erase the anisotropic effect of the
spanning cluster on the statistics. Otherwise, the probability $R_L^e$
(crossing at either directions) accentuate the anisotropy of
the percolating cluster.
Therefore, we do not expect a same result for $\nu$ using
$p_c(L)$ obtained from $R_L^e$ or $R_L^b$.

Figure \ref{fig5} shows $\nu$ versus $\rho$ for the two cases:
crossing in either directions (diamonds), and crossing in both
directions (triangles). We call $\nu^e$ and $\nu^b$ these two
exponents. The values of $\rho$ are indicated in the figure. Based
in this figure we initially
conclude: diversly from  $d_f$ and $\beta$,
the exponent $\nu$ depends on $\rho$.
 For the case $\rho=1$ (the square lattice)
 we have $\nu^e=1.33 $ and  $\nu^b= 1.39$, considering the
sensitivity of the method, these values are in good agreement
to the exact value of $\nu=\frac{4}{3}=1.333$.

  An anisotropic percolating cluster has a larger spanning probability
in either direction, $R_L^e$, than in both directions, $R_L^b$.
Therefore, as the anisotropy of the cluster increases, the
difference between the two estimations becomes more accentuated. This
result is observed in figure \ref{fig5}. The same result can
be verified using  $\Delta \nu = | \nu^e - \nu^b |$ which
increases as $\rho \rightarrow 0$.
In Table $II$ we show the data of figure \ref{fig5} and  $\Delta \nu$.

\centerline{\it Table II}
\begin{tabbing}
  $(s,r)$      \hspace{0.6cm}        \= (1,1) \hspace{0.7cm} \= (4,3) \hspace{0.7cm}\=      (3,2) \hspace{0.7cm} \= (2,1) \hspace{0.7cm} \= (5,2)     \hspace{0.7cm} \=(3,1)      \hspace{0.7cm} \= (4,1)   \hspace{0.7cm} \= (5,1)\\
\hspace{0.2cm} $\nu^{e}$  \hspace{0.9cm} \= 1.33 \hspace{0.8cm} \= 3.80  \hspace{0.8cm} \= 3.25  \hspace{0.8cm} \= 2.49   \hspace{0.8cm} \=  4.78   \hspace{0.8cm} \= 3.48  \hspace{0.8cm} \= 4.92   \hspace{0.8cm} \= 6.13 \\
\hspace{0.2cm} $\nu^{b}$  \hspace{0.9cm} \= 1.39 \hspace{0.8cm} \= 3.84 \hspace{0.8cm} \= 3.16  \hspace{0.8cm}  \= 2.03 \hspace{0.8cm} \= 3.53    \hspace{0.8cm} \= 2.65  \hspace{0.8cm} \= 2.98   \hspace{0.8cm} \= 3.41 \\
\hspace{0.0cm} $ \Delta \nu $ \hspace{0.7cm} \=0.06 \hspace{0.8cm} \= 0.04 \hspace{0.8cm} \= 0.09   \hspace{0.8cm}  \= 0.46  \hspace{0.8cm} \= 1.25   \hspace{0.8cm} \= 0.83    \hspace{0.8cm} \= 1.94   \hspace{0.8cm} \= 2.72 \\
 \end{tabbing}

 It is also observed in figure \ref{fig5} that generically $\nu$ decreases
with $\rho$. This phenomenon is related with the stretching effect that
increases as $\rho \rightarrow 0$. We have discussed this point in
section $\S 2$ in connection with the stretching of the blocks that forms
the tilling of the multifractal.

\section{4 - Final Remarks}

To summarize we study critical exponents and fractal dimensions of the
incipient infinite cluster constructed on a
self-affine multifractal $Q_{mf}$.
The quantities $\beta$ and $d_f$ do not depend on the local  symmetries of the
percolating cluster. As a consequence
these two numbers are equal to the ones of
the standard (isotropic) two dimensional percolation. On the other side,
the exponent $\nu$, which is related to the correlation length,
depends on the local symmetry properties of the percolation cluster.
 There are some points to consider in the
symmetry properties of $Q_{mf}$ and its isotropy.
The first is related with the stretching of
the blocks of $Q_{mf}$ which increases as $\rho \rightarrow 0$. Moreover
there is the topology of $Q_{mf}$, specially its coordination number, that
changes along the object.

We have in mind in the construction of the multifractal object
the modeling of fluid flow in geologic medium.
The multifractal pattern of $Q_{mf}$ is a candidate to
 describe petroleum reservoir
heterogeneities \cite{Hermann,Muller1},
 complex geological structures \cite{Muller2}
and to study  other geophysical situations
\cite{ Riedi, Hubert,Lovejoy}.

The anisotropy of the percolation cluster is revealed
 when we compare the percolation threshold and $\nu$ in either
direction and  in both directions. In fact,
 $R_L^b$ performs an average over both directions erasing
the anisotropic effect of the percolation cluster.
Moreover we estimate two $\nu$ exponents, $\nu^e$ and $\nu^b$,
which are associated with $R_L^e$ and  $R_L^b$. Typically the
correlation length and the exponents $\nu$ are larger for the
percolation over $Q_{mf}$ than in usual percolation. This fact is related to the
very stretched blocks in the multifractal that can make long range
connections in the object. We see that the $\nu$ exponent is able to
probe the local anisotropy of the percolation cluster on a
multifractal support. The value of $\nu$ that is identical to the
standard percolation ($\rho=1$, the symmetric square lattice) unfolds
into different values indicating different degrees of local
anisotropies.

To conclude, we estimate some quantities for percolation on the
multifractal object $Q_{mf}$ immersed in two dimensions. The exponent $\beta$ and
the fractal dimension of the percolation cluster are the same of standard
percolation in two dimensions. The exponent $\nu$ that is related with
the correlation length is influenced by the anisotropy of $Q_{mf}$. Depending
on the parameters that characterize the stretching of the tilling of $Q_{mf}$,
$\rho$, several values of $\nu$ appear. In this way the universality
class of percolation on $Q_{mf}$ depends also on $\rho$, the anisotropy of the tilling of the
multifractal.  This surprising result reveals that in this problem the local
symmetry of the support governed by $\rho$ affects strongly the critical percolation
behavior, leading to non-universality.

\vspace{2cm}

The authors gratefully acknowledge the financial support of Conselho Nacional
de Desenvolvimento Cient{\'\i}fico e Tecnol{\'o}gico (CNPq)-Brazil,
FINEP and CTPETRO.

\hspace{1cm}

\centerline{FIGURE LEGENDS}

\begin{figure}[ht]
\begin{center}
\caption{ The two initial steps in the formation of $Q_{mf}$.
In (a) a vertical line cut the
square in two pieces according to the ratio $\rho=\frac{2}{3}$.
Two horizontal lines cutting the rectangles by
 the same ratio are depicted in (b).
The second step is show in (c) and (d), the same process
of step $1$ is repeated for each former block. }
\label{fig1}
\end{center}
\end{figure}
\begin{figure}[ht]
\begin{center}
\caption{The object $Q_{mf}$ for the same
 $\rho=\frac{2}{3}$ used in figure \ref{fig1},
but the number of steps is evolved until $n=8$.
Figure (b) shows a zoom of the internal square of figure (a). }
\label{fig2}
\end{center}
\end{figure}
\begin{figure}[ht]
\begin{center}
\caption{A view of the percolating cluster for one
typical realization of $\rho=\frac{1}{3}$.
}
\label{fig3}
\end{center}
\end{figure}
\begin{figure}[ht]
\begin{center}
\caption{A graphic of  $ln \left( p_c(L) -p_c \right)$ against $ln L$ for
$\rho=\frac{1}{3}$.  It is indicated in the figure the
equation of linear regression of the data.
 The slope of the curve is the exponent $\nu$.}
\label{fig4}
\end{center}
\end{figure}
\begin{figure}[ht]
\begin{center}
\caption{A graphic of values of the  exponents
$\nu^e$ (square) and $\nu^b$ (circle) versus the anisotropy
parameter $\rho$.}
\label{fig5}
\end{center}
\end{figure}
\end{document}